# B PRODUCTION AT THE LHC / QCD ASPECTS

V. P. ANDREEV

*University of California at Los Angeles, Department of Physics and Astronomy,*

*475 Portola Plaza, Los Angeles, CA 90095-1547, USA*

On behalf of the ATLAS and CMS Collaborations

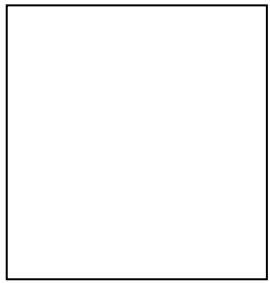

The LHC provides new opportunities to improve our understanding of the *b* quark using high statistics data samples and the 14 TeV center-of-mass energy. The prospects to measure the cross section for inclusive *b* production in events containing jets and at least one muon are presented. Studies of detector systematic effects and theoretical uncertainties are included. QCD aspects of the beauty production are discussed.

## 1 Inclusive b-quark production at LHC

### 1.1. *QCD aspects*

*B* production will be one of the most copious sources of hadrons at LHC. Three mechanisms contribute to the beauty production at hadron colliders: gluon-gluon fusion and $q\bar{q}$ annihilation (flavor creation in hard QCD scattering), flavor excitation (semi-hard process) and gluon splitting (soft process). It is important to measure the *B*-hadron $p_T$ spectra within large range to be able to disentangle the contributions of those mechanisms. Flavor creation refers to the lowest-order, two-to-two QCD $b\bar{b}$ production diagrams. Flavor excitation corresponds to diagrams where a $b\bar{b}$ pair from the quark sea of the proton is excited into the final state due to one of the *b* quarks undergoes a hard QCD interaction with a parton from the other proton. Gluon splitting refers to the processes in which the $b\bar{b}$ pair arises from a $g \to b\bar{b}$ splitting in the initial or final state. Neither of the quarks from $b\bar{b}$ pair participate in the hard QCD scattering in this case. Inclusive *b*-quark production has been studied at other proton and electron colliders. The observed shapes of distributions and correlations are reasonably well explained by perturbative QCD. However, the observed cross-sections at the Tevatron (Run I) are larger than QCD predictions [1-3] which is confirmed by Run II data. Similar effects are observed in γp collisions at HERA [4-6] and in γγ interactions at LEP [7,8]. The agreement between experiment and theory has improved due to more precise parton density functions and proper estimates of fragmentation effects [9]. But an agreement is not complete and phenomenological input to the calculations is required.

1.2. *B production measurement at LHC*

A study [10] has been performed in CMS on Monte Carlo events generated with PYTHIA [11] to investigate methods of identifying in CMS of *b*-jets (*b* "tagging") in an inclusive sample of events containing jets and at least one muon. We present here the capability to measure the inclusive *b*-quark production cross section as a function of the *B*-hadron transverse momentum $p_T$ and pseudorapidity η. The study of the CMS capability to measure the inclusive *b* production is based on full detector simulation. The measurement of the differential cross sections is studied for *B*-hadrons of $p_T$ > 50 GeV/c and within the fiducial volume of |η| < 2.4. The event selection requires a *b*-tagged jet in the fiducial volume to be present in the event. *B* tagging is based on inclusive secondary vertex reconstruction in jets [12]. At Level-1 (L1) trigger, the single muon trigger is used. At the High Level Trigger (HLT) we require the "muon + *b*-jet" trigger. The most energetic *B*-hadron inside the phase space defined above is selected. Good correspondence between the generated *B*-particle and the reconstructed *b*-tagged jet is observed. The corresponding relative resolutions for *B*-particles with $p_T$ > 170 GeV/c are 13% and 6% for $p_T$ and pseudorapidity, respectively. The average *b* tagging efficiency is 65% in the barrel region, while the efficiency is about 10 % less for the endcap region.

The signal fraction is determined from a fit to the data distribution using the simulated shapes for the signal and background. To do so we apply a lepton tag by selecting inclusive muons. Each reconstructed muon is associated to the most energetic *b* tagged jet. The muon must be closer to this *b* tagged jet than to any other jet in the event. Otherwise the event is discarded. In most cases the tagged muon is inside the b jet. The average efficiency of associating the muon with the b tagged jet is 75 %. We calculate the transverse momentum of the muon with respect to the *b*-jet axis which effectively discriminates between *b* events and background. Figure 1 shows an example of the fit of the distribution of the muon $p_T$ with respect to the closest jet, using the expected shapes for the muons from *b* events, charm events and light quark events. The normalisation of the three contributions are free parameters in the fit. The event fractions are well reproduced within statistical errors [10].

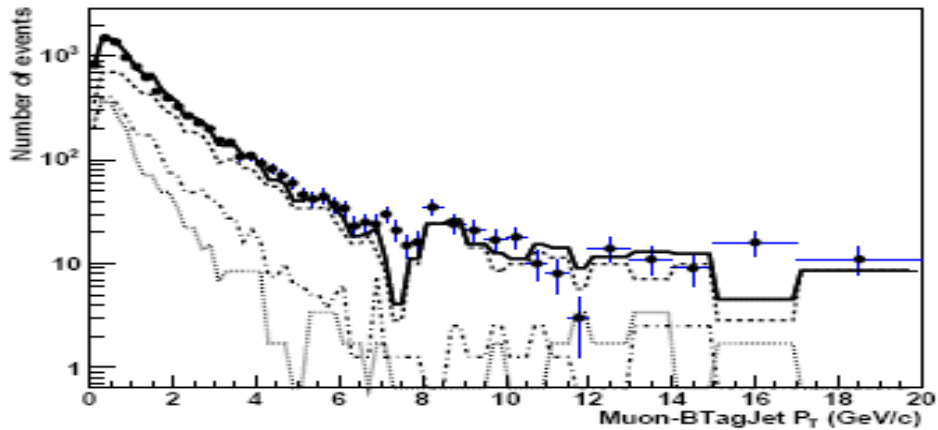

Fig. 1. Fit of the muon $p_T$ spectrum with respect to the closest *b* tagged jet. The contributions of tagged muons from *b* events (dashed curve), *c* events (dot-dashed curve) and light quark events (dotted curve) as defined by the fit are shown. The solid curve is the sum of the three contributions.

The total event selection efficiency is about 5 %. By correcting for the semi-leptonic branching ratio of $b$ quarks and $c$ quarks it amounts to about 25 % on average. It turns out that the total efficiency is almost independent of transverse momentum and angle of the $B$-particle. Therefore the measurement of the differential cross section is less affected by systematic uncertainties due to bin-by-bin efficiency corrections.

Several potential sources for systematic uncertainties are considered and their impact on the observed cross section is detailed in [10]. The largest uncertainty arises from the 3 % error [13] on the jet energy scale which leads to a cross section error of 12 % at $E_T > 50$ GeV/c. The estimated statistical, systematic and total uncertainty as function of the $b$ tagged jet transverse momentum with respect to the beam line is shown in Figure 2.

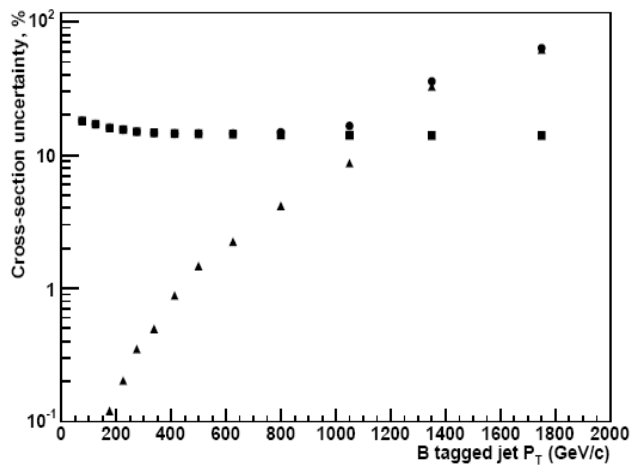

Fig. 2. The statistical uncertainty for the cross section measurement (triangles), systematic (squares) uncertainty and total (dots) uncertainty as function of the b tagged jet transverse momentum with respect to the beam line. Total uncertainty comprises the statistical and systematic uncertainties added in quadrature.

The event selection for inclusive $b$ production measurement at CMS will allow to study $b$ production mechanisms on an event sample of 16 million $b$ events for 10 fb$^{-1}$ of integrated luminosity. The $b$ purity of the selected events varies as function of the transverse momentum in a range from 70 % to 55 %. Our estimate shows that with the CMS detector we can reach 1.5 TeV/c as the highest measured transverse momentum of $B$ hadrons. The results are preliminary, the improvements are likely as further jet calibration tunings, software and analysis algorithm developments are foreseen.

1.3. $b\bar{b}$ correlations

Correlations measurements are foreseen at LHC in order to study details of the production mechanisms discussed in the chapter 1.1. The angular distance $\Delta\varphi$ between the $b$ quark directions in the transverse plane is the main discriminating variable to disentangle contributions from the gluon-gluon fusion, gluon splitting and flavor excitation. The $\Delta\varphi$ distribution for gluon splitting is slightly peaked at small $\Delta\varphi$ values. The angle between the two $b$-quarks produced by the gluon-fusion mechanism has a peak at 180 degrees, as expected, since in the process $gg \to b\bar{b}$ the $b$ quarks are produced back-to-back in the transverse plane. For the flavor excitation production mechanism the back-to-back topology is preferred too. The MC study with the CMS detector simulation are presented in [14].

ATLAS collaboration is going to use $J/\Psi$ from decay of one $B$ particle and semi-leptonic muon decay from another $B$ for the $b\bar{b}$-correlations measurement [15]. The $\Delta\varphi$ between $J/\Psi$ and muon directions distribution is shown in Figure 3. The muon efficiency as function of $\Delta\Phi_{J/\Psi-\mu}$ is also plotted. The efficiency is rather flat function, which will allow to avoid any significant distortion of the original spectrum.

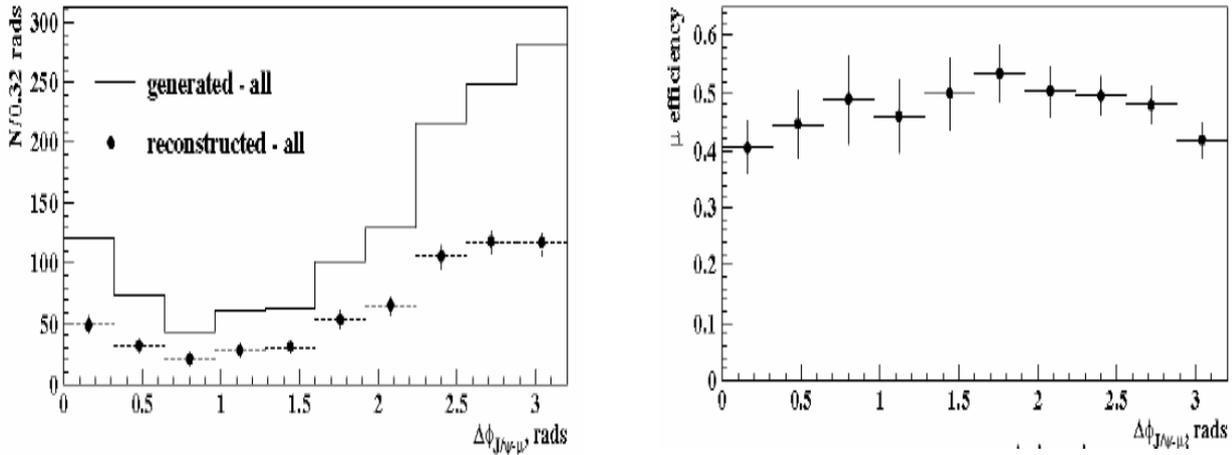

Fig. 3. ATLAS study for the $b\bar{b}$-correlations measurement. The left plot shows the $\Delta\varphi$ between $J/\Psi$ and muon directions at generator level and after the reconstruction. The right plot shows muon efficiency as function of $\Delta\varphi$.

## 2  Conclusion

All LHC experiments will measure *B* production cross section at the new accelerator energy frontier, 14 TeV center-of-mass energy. This will provide a new test of the QCD and will fix the normalization for the beauty events background in the new physics searches.